\documentclass[apj]{emulateapj}
\pdfoutput=1

\usepackage{apjfonts}

\usepackage{graphicx}
\usepackage{amsbsy}
\usepackage{mathrsfs}
\usepackage{mathpazo,bm}
\usepackage{natbib}

\newlength{\hfwidth}
\newlength{\hfwidthsingle}
\newlength{\figspace}
\newcommand{\beq}{\begin{equation}}
\newcommand{\eeq}{\end{equation}}
\newcommand{\pderiv}[2]{\frac{\partial{#1}}{\partial{#2}}}

\newcommand{\aderiv}[1]{\frac{\mathrm{D}{#1}}{\mathrm{D}t}}

\newcommand{\xtimes}[2]{#1\times{10^{#2}}}

\renewcommand{\v}[1]{{\boldsymbol{#1}}} 

\newcommand{\ksi}{\xi}

\newcommand{\del}{\v{\nabla}}
\newcommand{\grad}{\del}
\newcommand{\Div}{\del\cdot}

\newcommand{\cv}{c_{_{V}}}

\newcommand{\cs}{c_\mathrm{s}}

\newcommand{\Eq}[1]{Eq.~(\ref{#1})}

\newcommand{\eq}[1]{\Eq{#1}}

\newcommand{\Fig}[1]{Fig.~\ref{#1}}
\newcommand{\fig}[1]{\Fig{#1}}

\newcommand{\sect}[1]{Sect.~\ref{#1}}

\shorttitle{Buoyancy-unstable planetary wake}
\shortauthors{Lyra et al.}

\slugcomment{Draft version}

\begin{document}

\title{On shocks driven by high-mass planets in radiatively
  inefficient disks. II. Three-dimensional global disk simulations.}
\author{Wladimir Lyra\altaffilmark{1,2,3}, Alexander
  J.W. Richert\altaffilmark{4}, Aaron Boley\altaffilmark{5}, \\
  Neal Turner\altaffilmark{2}, Mordecai-Mark Mac Low\altaffilmark{6,7},\\
  Satoshi Okuzumi\altaffilmark{2,8}, and Mario Flock\altaffilmark{2}}

\altaffiltext{1}{Department of Physics and Astronomy, California State
  University Northridge, 18111 Nordhoff St, Northridge, CA 91330, USA. wlyra@csun.edu}
\altaffiltext{2}{Jet Propulsion Laboratory, California Institute of Technology, 4800 Oak Grove Drive, Pasadena, CA, 91109, USA. neal.j.turner@jpl.nasa.gov,mario.flock@jpl.nasa.gov}
\altaffiltext{3}{Division of Geological \& Planetary Sciences, California Institute of Technology, 1200 E California Blvd MC 150-21, Pasadena, CA 91125 USA}
\altaffiltext{4}{Department of Astronomy \& Astrophysics, Penn State University, 525 Davey Lab, University Park, PA 16802, USA. ajr327@psu.edu} 
\altaffiltext{5}{Department of Physics and Astronomy, University of British Columbia, 6224 Agricultural Road, Vancouver, BC V6T 1Z1, Canada. acboley@phas.ubc.ca}
\altaffiltext{6}{Department of Astrophysics, American Museum of
  Natural History, Central Park West at 79th Street, New York, NY
  10024-5192, USA; mordecai@amnh.org}
\altaffiltext{7}{Institut f\"ur Theoretische Astrophysik, Zentrum
  f\"ur Astronomie der Universit\"at Heidelberg,
  Albert-Ueberle-Str.~2, 69121 Heidelberg, Germany.}
\altaffiltext{8}{Department of Earth and Planetary Sciences, Tokyo Institute of Technology, 2-12-1 Ookayama, Meguro-ku, Tokyo 152-8551, Japan. okuzumi@geo.titech.ac.jp}

\begin{abstract}
Recent high-resolution, near-infrared images of protoplanetary disks
have shown that these disks often present spiral features. Spiral arms
are among the structures predicted decades ago by numerical simulations of 
disk-planet interaction and thus it is tempting to suspect that
planetary
perturbers are responsible for the observed signatures. However, such 
interpretation is not free of problems. The spirals are found to
have large pitch angles, and in at least one case (HD 100546) the spiral feature appears effectively
unpolarized, which implies thermal emission of the order of
  1000\,K ($465 \pm 40$\,K at closer inspection). We have recently shown in two-dimensional models that shock dissipation
in the supersonic wake of 
high-mass planets can lead to significant 
heating if the disk is sufficiently adiabatic. In this paper we
extend this analysis to three dimensions in thermodynamically evolving
disks. We use the {\sc Pencil Code} in spherical coordinates for
our models, with a prescription for thermal cooling
based on the optical depth of the local vertical gas column. 
We use a 5$M_J$ planet, and show that shocks in the
region around the planet where the Lindblad resonances occur
heat the gas to
substantially higher temperatures than the ambient disk gas at that
radius. The gas is accelerated vertically away from the
midplane by the shocks to form shock
bores, and the gas falling back toward the midplane breaks up into a
turbulent surf near the Lindblad resonances. This turbulence, although localized, has
high $\alpha$ values, reaching 0.05 in the inner Lindblad resonance, and 0.1 in
the outer one. We also find evidence that the 
disk regions heated up by the planetary shocks
eventually becomes superadiabatic, generating convection far from the planet's
orbit.  
\end{abstract}
\subjectheadings{hydrodynamics --- planet-disk interactions ---
  planets and satellites: formation --- protoplanetary disks --- shock
  waves --- turbulence}

\section{Introduction}
\label{sect:introduction}

Recent observations of protoplanetary disks reveal clear spiral
features (e.g., Muto et al.\ 2012, Garufi et al.\ 2013, Currie et
al. 2014, Benisty et al.\ 2015), as seen in polarized scattered light
in near infrared. Given the high density of the disk, the micron-sized dust grains responsible for the scattering must be
suspended high in the disk atmosphere to produce these features. 

Numerical simulations of planet-disk interaction (e.g. Bryden et
al. 1999, Kley 1999, Lubow et al.\ 1999, Baruteau et al. 2014) have shown
the formation of distinctive one-armed spirals. Even the very first
of such calculations (Miki 1982), 
although done in two dimensions (2D) in a local box, already shows evidence of a spiral
pattern excited by the
planet. High-mass planets can lead to strong spiral shocks,
  while low-mass planets (e.g. Neptune) produce spirals with only
  slight density enhancements. In the low-mass limit, the perturbations can be treated using linear
theory, and Ogilvie \& Lubow (2002) showed that the 
shape of the spiral density perturbation launched by such a low-mass planet 
can be predicted analytically as a superposition of the individual 
Fourier modes.

Although it is tempting to assume that the observed spiral
features are indeed due to planets, several of the observed spirals
have pitch angles too wide to fit the analytical
predictions based on linear theory in the simplest case of an azimuthally-isothermal
disk. Since the pitch angle in the linear theory depends on
the sound speed, this implies that the spirals are launched
from higher temperatures than the ambient gas. MWC 758, for instance
has a measured background temperature of
$\approx$53K at 50\,AU (Isella et al.\ 2010), yet a fit to the
analytical model requires a temperature of
$\approx$300K to match the observed spirals (Benisty et al.\ 2015). 

Nonetheless, if planets do produce these spirals, the simplest
  explanation for the observed spiral morphologies may be that the
  planets are too massive for their influence on the disk to be
  described by linear theory. The linear theory requires the planets to be of
low mass, so that the wake launched is subsonic. If the planet is
massive, the wake launched propagates faster than the speed of sound,
and steepens into a shock near the planet. The shape of
the spiral shock then rapidly deviates from that predicted from linear
theory. 

Evidence of this behavior has been abundantly clear in the literature
for at least a decade, visibly seen in the disk-planet simulations in
the code comparison of de Val-Borro et al.\ (2006). In that
study, 14 independent codes were used to reproduce viscous and
inviscid simulations of Neptune-mass and Jupiter-mass planets. Whereas
the shape of the spirals launched by the Neptune-mass planets were
described well by linear theory (Ogilvie \& Lubow 2002), the spirals launched
by the Jupiter-mass planets deviated from it substantially. It was
understood that this feature was due to the Jupiter-mass planet driving
shocks that were outside the range of applicability of linear
theory. Dong et al. (2015) and Zhu et al.\ (2015) have recently shown that a weakly nonlinear
theory (Goodman \& Rafikov 2001, Rafikov 2002) can reproduce the
pitch angle of spiral shocks launched by high-mass planets. 

Yet, even though a higher pitch angle 
can reproduce the spiral launched by high-mass planets reasonably
  well, without requiring {\it ad hoc} high 
aspect ratios, there are still features that require high
temperatures. For instance, in the circumstellar material 
around HD 100546, a spiral-like feature is seen (Currie et al.\ 2014, 2015) 
showing little polarization. This implies that the emission is thermal 
rather than scattered starlight, and thus must of the order of 
1000 K to reproduce the infrared brightness (in
  \sect{sect:conclusions} we examine the observational data and determine
the temperature needed to reproduce the H and L$^\prime$ magnitudes to
be 465$\pm$40 K).

We have recently shown in
Richert et al.\ (2015, hereafter Paper I) that when 
a planet is massive enough to produce a spiral wave that
steepens into a shock, the associated shock dissipation can
significantly heat the disk.  If the radiative efficiency of the disk
is low, the temperature can rise until buoyancy upsets the force
balance. This can cause disk turbulence with significant angular
momentum transport.  Effectively, the gravitational potential well of the planet powers 
a vigorous heat source. 

Here we report three-dimensional (3D) simulations exploring the same
behavior. There are significant qualitative differences between 2D and
3D in this problem. The midplane is close to adiabatic, but the
atmosphere of the disk behaves more isothermally because it can more
quickly cool radiatively. This effect alone will
already lead to a different behavior in 3D, since now the energy buildup due
to shocks has to occur within the time that it takes for energy 
to diffuse from the midplane to the atmosphere (from which it 
escapes efficiently). Also, upward
motions that bring a gas parcel from the midplane into the atmosphere
will quickly lead to efficient cooling. Another major difference is that a shock will be
weaker in 3D, because the extra dimension translates into an extra degree of
freedom. Therefore, when the shock expands adiabatically, it can accelerate material in
the vertical direction. This was discussed in detail in Boley \&
Durisen (2006), who referred to the structure as shock bores,
following the work of Gomez \& Cox (2004) on galactic disk shocks. 
As gas shocks and
accelerates upwards, a column of hot, thin gas forms around the colder
and denser gas just outside of the spiral shocks.  When the jetted
material falls back towards the disk surface, breaking waves can form,
similar in appearance to beach surf.  If strong enough, these waves should produce
localized turbulence around the locations of the Lindblad resonance.  
We intend in this paper to find signatures of this process in
3D, non-isothermal disks containing high-mass planets. 

Although our work resembles that of Zhu et al.\ (2015) in that
  we both model 3D, nonisothermal disks with high-mass planets in
  order to examine the effect of their shocks, there are also
  significant differences in technique and focus. While Zhu et al.\
  (2015) use a fixed cooling time with an exponential fall to
  isothermal beyond three scale heights, we use a
  dynamical cooling time dependent on vertical optical depth, that varies following the movements
  of the material.  Zhu et al.\ (2015) examined the
  shape of the spiral launched by the planet, while we focus on the
  effect of the shocks on disk accretion and energy budget. Because of that, Zhu
  et al.\ only ran their disks for 10 orbits, halting before gap
  opening and substantial alteration of the disk structure, while we
  ran to 40 orbits, long enough to achieve gap opening.

This paper is structured as follows. In Sect~2 we describe the model
used. The results are presented in Sect~3, followed by discussions and
conclusions in Sect~4.

\begin{figure}
  \begin{center}
    \resizebox{\columnwidth}{!}{\includegraphics{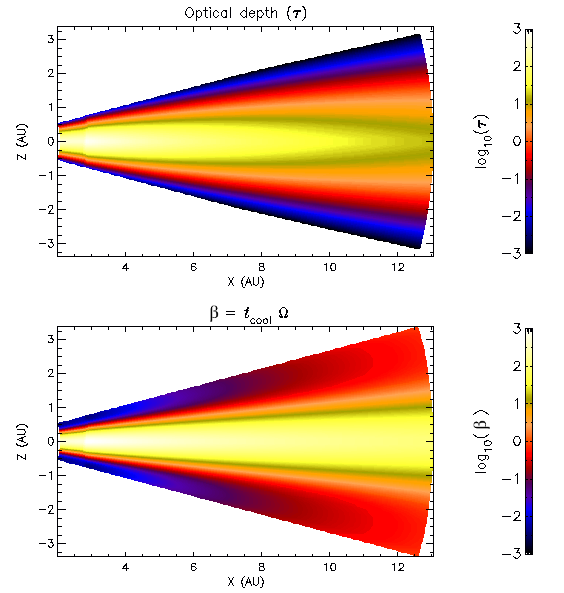}}
  \end{center}
  \caption[]{Optical depths and cooling times of our initial disk. The midplane is
    nearly adiabatic, while the upper layers are nearly
    isothermal. Calculated using the opacities of Bell et al.\ (1997), and
    \eq{eq:tcool}. The dimensionless cooling time is $\beta=t_{\rm cool}
    \varOmega_K$. }
  \label{fig:opacity-cooling}
\end{figure}

\section{The model}

\label{sect:model}
\subsection{Model equations}

\begin{figure*}
\begin{center}
  \resizebox{.545\textwidth}{!}{\includegraphics{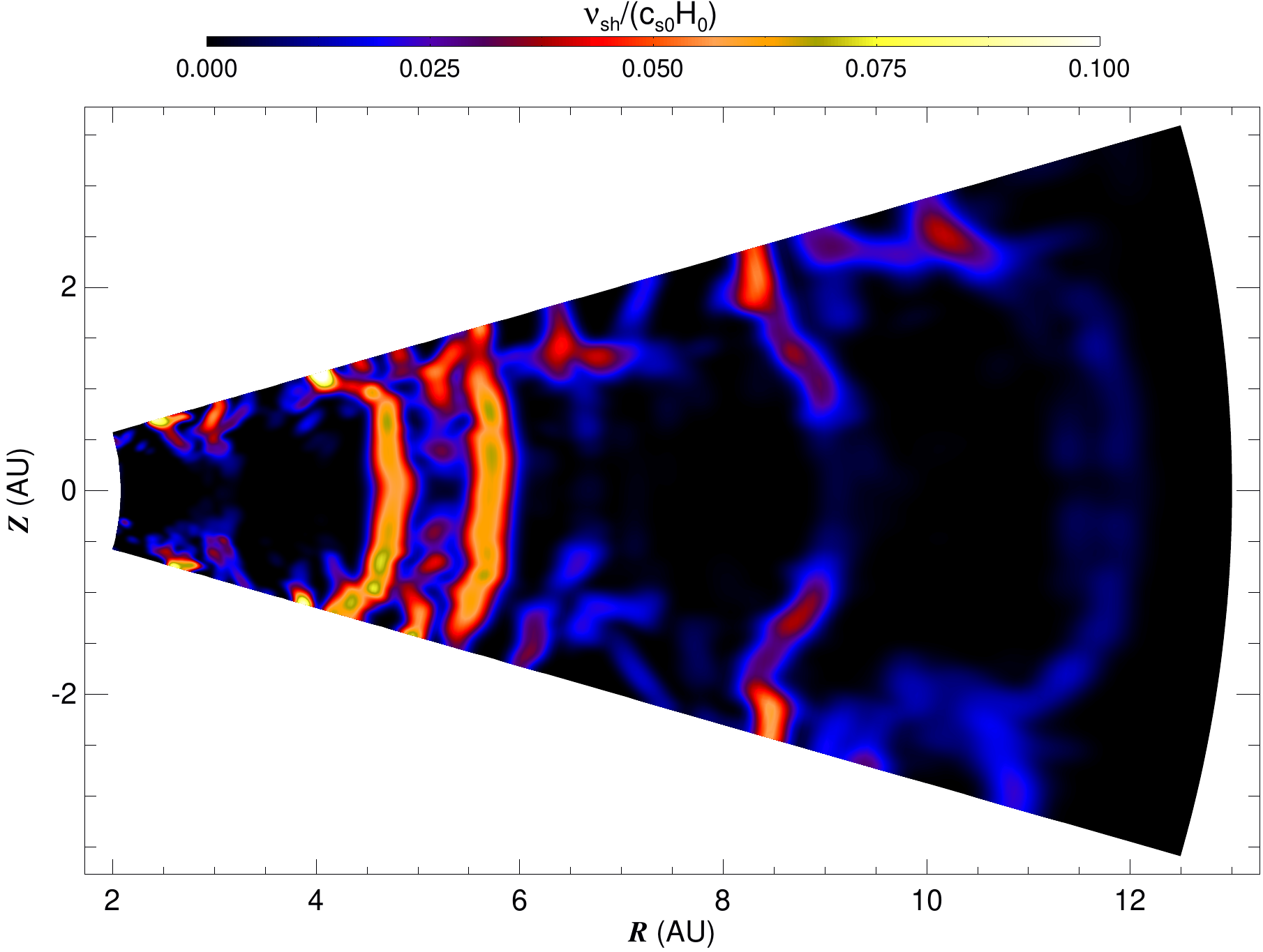}}
  \resizebox{.445\textwidth}{!}{\includegraphics{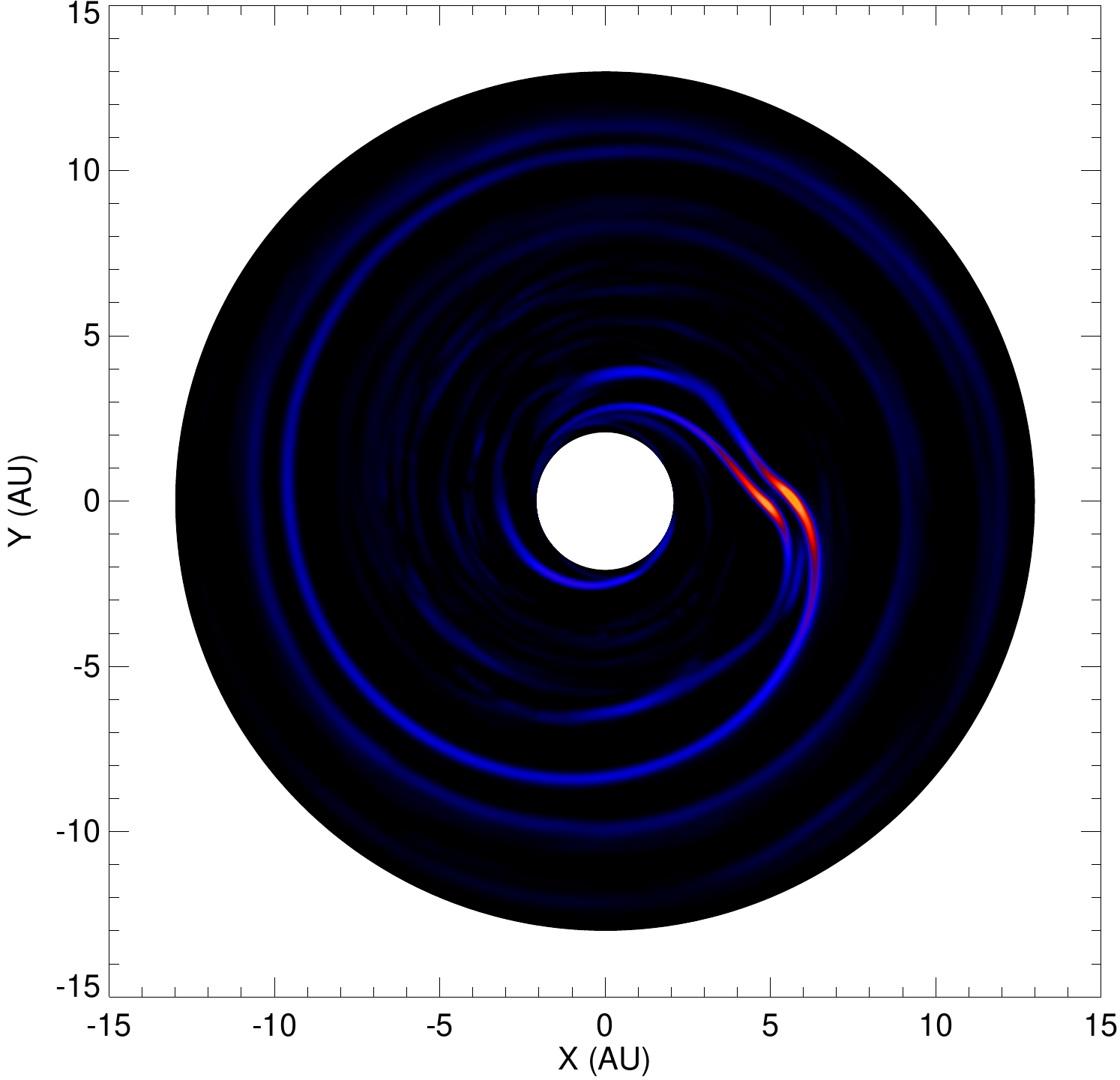}}
\end{center}
\caption[]{The location of shocks as traced by the shock viscosity \eq{eq:shock-visc} normalized by $c_sH$, in the
  vertical (meridional) plane that contains the planet (left), and in the disk midplane
  (right). The shock viscosity is a measure of the positive
    part of $\Div{\v{u}}$, so it only becomes significant in
  locations where shocks occur. The planet is
  located at 5.2\,AU. Vertical shock bores are visible in the
  meridional plane, extending well into the disk atmosphere. In the
  midplane plot (right), the shocks are seen to be associated with
  the spiral the planet launches from its Lindblad resonances.}
\label{fig:shocks}
\end{figure*}

\begin{figure}
  \begin{center}
    \resizebox{\columnwidth}{!}{\includegraphics{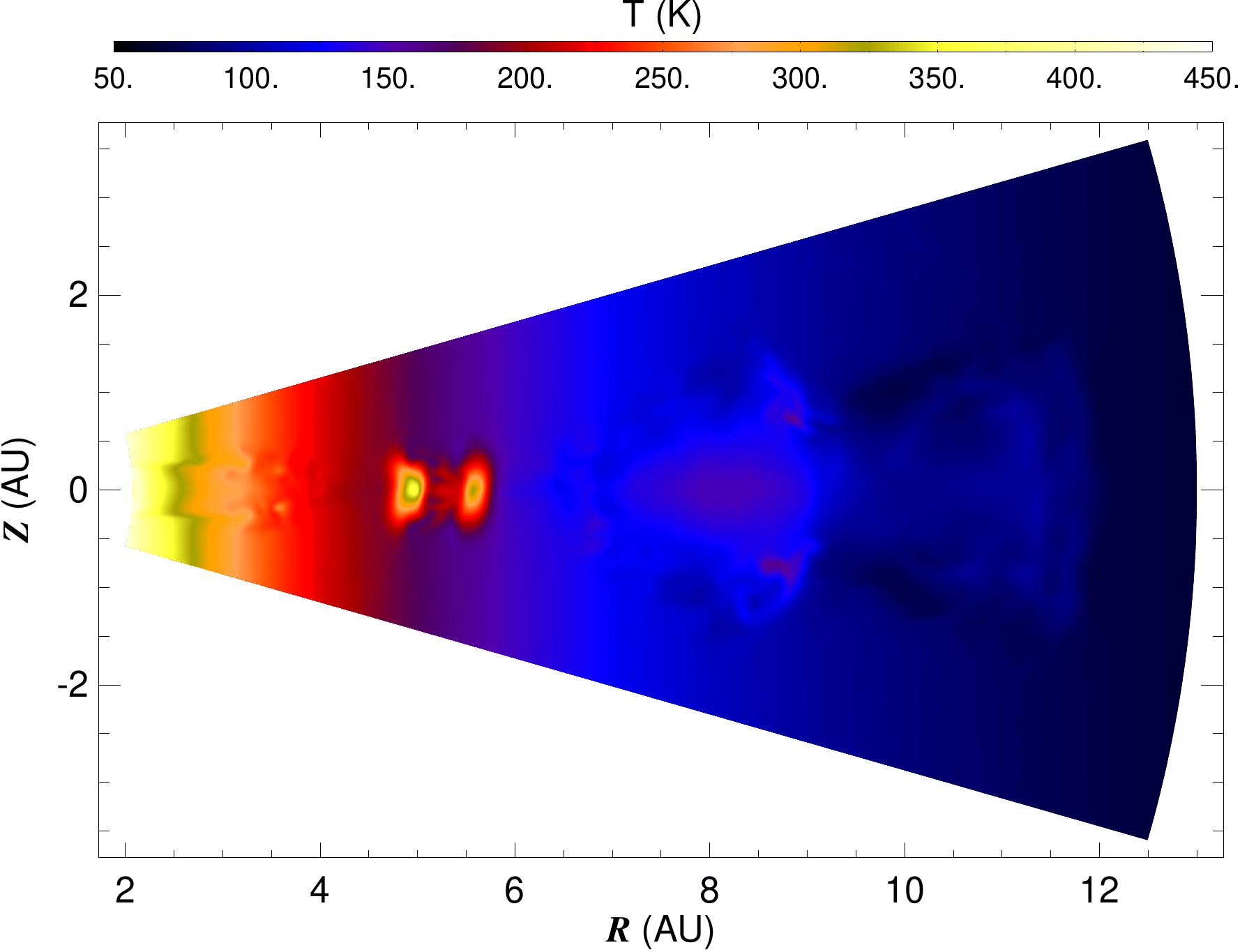}}
  \end{center}
  \caption[]{Temperature structure in the vertical plane containing
    the planet. Where the shock 
    bore intersects the more radiatively inefficient regions of the
    disk, the temperature rises considerably.}
  \label{fig:temperature-meridional}
\end{figure}

The models shown in this paper were calculated with the {\sc Pencil
  Code}\footnote{The code, with improvements done for this work, is
  publicly available at \url{http://pencil-code.nordita.org}.}.  We
solve the equations of hydrodynamics in spherical coordinates in the reference frame with
the origin at the star and following the motion of the planet. In this frame,
both the star and the planet are stationary. The equations of motion
are 
  
\begin{eqnarray}
  \aderiv{\ln \rho} &=& -\Div\v{u}, \label{eq:continuity}\\
  \aderiv{\v{u}} &=& -\frac{1}{\rho}\grad p - \grad \varPhi +
                     \frac{1}{\rho}\Div{\v{\zeta}}\nonumber \\ 
                & &-2\v{\varOmega}_p \times \v{u}  - \v{\varOmega}_p \times (\v{\varOmega}_p \times \v{r}) ,  \label{eq:navier-stokes}\\
  T  \aderiv{s} &=& - \cv \frac{\left(T-T_{\rm ref}\right)}{t_{\rm cool}} + \varGamma_{\rm sh},\label{eq:entropy-equation}
\end{eqnarray}

In these equations, $\rho$ is the density, $t$ the time, $\v{u}$ 
the velocity, $p$ the pressure, $\varPhi$ the gravitational
potential, $\v{\zeta}$ the shock viscosity tensor, $\v{\varOmega_p}$ the
angular velocity vector of the planet, $\v{r}$ the distance to the
origin, $s$  the entropy, $T$ temperature, $\cv$ the specific
heat at constant volume, $T_{\rm ref}$ a reference temperature,
$t_{\rm cool}$ the cooling time, and $\varGamma_{\rm sh}$ the shock
heating. The advective derivative 

\begin{equation}
  \aderiv{} \equiv \pderiv{ }{t} + \v{u}\cdot\del.
\end{equation} 

\noindent The gravitational
potential 

\begin{equation}
  \varPhi = - \frac{GM_\star}{r} -\frac{GM_p}{(|\v{r}-\v{r}_p|^2 + b^2)^{1/2}} + \frac{GM_p}{r_p^3} \v{r} \cdot \v{r}_p
\end{equation}

\noindent includes the contribution of the star (first term), planet
(second term), and the indirect terms due to the motion of the star
(third term). Here, $G$ is the gravitational constant, $M_\star$ the
star mass, $M_p$ the planet mass, and $\v{r}_p=r_p\hat{\v{x}}$ the vector position of
the planet. The softening radius $b$ in the planet potential is included to avoid
singularities, and is set to the planet's Hill radius. Note that
the planet's orbital frequency

\begin{equation}
  \v{\varOmega}_p=\left(\frac{G(M_\star+M_p)}{r_p^3}\right)^{1/2} \v{\hat{z}}.
\end{equation}

In Pencil, we capture shocks using an explicit shock viscosity
  prescription as first proposed by
  von Neumann \& Richtmyer (1950). The third term in the momentum equation (\eq{eq:navier-stokes})
is the viscosity required to spread shocks out to resolvable width,
and the last term in the entropy equation (\eq{eq:entropy-equation}) is the shock viscous
heating. 

The shock viscosity tensor takes the form of a bulk viscosity

\begin{equation}
  \v{\zeta}_{ij}= \nu_{\rm sh} \, \rho \, \delta_{ij} \ \Div\v{u},
  \label{eq:zeta}
\end{equation}

\noindent and the associated shock heating is 

\begin{equation}
  \varGamma_{\rm sh} = \nu_{\rm sh} \left(\Div\v{u}\right)^2.
\label{eq:shocktemp}
\end{equation}

\noindent They depend on the shock viscosity, which we
  define numerically as

\begin{equation}
  \nu_{\rm sh } = c_{\rm sh}
  \left<\max_3[(-\Div\v{u})^+]\right>{\left[\min(\Delta x)\right]}^2{.}
  \label{eq:shock-visc}
\end{equation}

\noindent The actual form of the shock viscosity that we use has
  been described in Haugen et al. (2004) and in paper I. The superscript plus sign
  indicates the positive part of the quantity. In paper I we experimented with a range of values for the
  coefficient $c_{\rm sh}$, and found that it did not affect the outcome of
  the simulations. As long as the shock is resolved, the value of the
  shock viscosity coefficient does not change the amount of heating;
  rather, it just changes the volume (number of grid cells) over which the
  shock energy is spread.

We use for equation of state 

\begin{equation}
  p = \rho \cs^2/\gamma,
\end{equation}

\noindent where $c_s$ is the adiabatic sound speed, related to the temperature by 

\begin{equation}
  T=\cs^2/[c_p(\gamma-1)],
\end{equation} 

\noindent $c_p=\gamma\cv$ is the specific heat at constant
pressure, and $\gamma$ the adiabatic index. The specific heats are
related to the universal gas constant $\mathcal{R}$
by 

\begin{equation}
\mathcal{R}=\mu(c_p-\cv),
\end{equation} 

\noindent where $\mu$ is the mean molecular weight of the gas.

In the entropy equation, we use Newton cooling with a dynamical
cooling time $t_{\rm cool}$ that varies following the movements
  of the material.  We define $t_{\rm cool}$ based on the radiative timescale
$t_{\rm rad}=E/\dot{E}$, where $E=\cv\rho T$ is the gas internal energy. 
Equating $\dot{E}=\Div{\v{F}}$ and integrating to remove the divergence, we define 

\begin{equation}
  t_{\rm cool} \equiv \frac{\int E dV}{\int F \v{\hat{n}}\cdot d\v{A} }.
  \label{eq:tcool-def}
\end{equation}

\noindent Here, the flux $F=|\v{F}|=\sigma T^4/\tau_{\rm eff}$ and the effective optical
depth, accounting for both the opaque and transparent limits,
is given by (Hubeny 1990, D'Angelo et al. 2003)

\begin{equation}
  \tau_{\rm eff} = \frac{3\tau}{8} + \frac{\sqrt{3}}{4} + \frac{1}{4\tau}.
\end{equation}

For the integration we assume a Gaussian sphere of radius equal to the
scale height, the largest isotropic scale

\begin{equation}
  H=\frac{1}{\sqrt{\gamma}}\frac{c_s}{\varOmega_K},
\end{equation}

\noindent where 

\begin{equation}
\varOmega_K^2=\frac{G(M_\star+M_p)}{R^3}.
\end{equation}

\noindent is the Keplerian frequency. We assume as an approximation that $E$ and $F$
are constant in that sphere and equal to the value at the center, and thus write the cooling time as 

\begin{equation}
  t_{\rm cool} =  \frac{\cv \rho  H \tau_{\rm eff}}{3\sigma  T^3 }.
  \label{eq:tcool}
\end{equation}

This cooling prescription was used in Lyra et al. (2010) and
  Horn et al. (2012) for one-dimensional vertically averaged disk
  models. Averaging instead over cylinders or over horizontal layers
  yields different geometric factors (2 and 4/3, respectively, where
  we have 3; Horn et al. 2012, D'Angelo et al. 2003). The range among
  the geometrical factors is considerably less than the range in
  opacity allowed by dust abundance variations, so any of the three
  approaches could safely be used.

The dimensionless cooling time is defined as 

\begin{equation}
  \beta \equiv  t_{\rm cool}\varOmega_K.
\end{equation}

The optical depth $\tau$ is the vertical integral of $\rho \kappa$
where $\kappa$ is the opacity, taken from Bell et al.\ (1997), updated
by Semenov et al.\ (2004). For every grid cell, we calculate the optical depth due to
material above and below

\begin{eqnarray}
\tau_{_{\rm upper}} &=& - \int_{\infty}^{z} \rho \kappa dz^\prime\\
\tau_{_{\rm lower}} &=&    \int_{-\infty}^{z} \rho \kappa dz^\prime
\end{eqnarray}

\noindent and take $\tau = {\rm min}(\tau_{_{\rm upper}},\tau_{_{\rm lower}})$. In
practice, instead of integrating in $z$ we simply integrate in
$\theta$, which simplifies the amount of required communication in a spherical
grid. Since the disk is geometrically thin, and the departure between surfaces
of constant $z$ and surfaces of constant $\theta$ is more pronounced only for
the optically thin, high-$z$ material, the error introduced is
small. The reference temperature $T_{\rm ref}$ is set to the initial
temperature at every radius.

Sixth-order hyper-dissipation terms are added to the evolution equations to
provide extra dissipation near the grid scale, as discussed in Lyra et
al. (2008). These terms are needed for
  numerical stability because the high-order scheme of the
Pencil Code has little overall numerical dissipation (McNally et
al. 2012). 
They are chosen to produce Reynolds numbers of order unity at the grid
scale, but then drop as the sixth power of the scale at larger scales, so
that they have negligible influence on the large-scale flow.

\subsection{Initial conditions}

We model the disk in spherical coordinates, using a logarithmic grid
in the radial direction that samples a constant number of grid cells
per scale height (in the radial direction). The grid is
    uniform in the meridional and azimuthal directions, and has resolution
($N_r,N_\theta,N_\phi$)=(256, 128, 768). The radial range is
[0.4,2.5], the meridional range spans $4H$ above and below the midplane,
which corresponds to [-0.28,0.28] radians ($\approx 16^\circ$ to each side). The azimuthal range is $2\pi$. 

The temperature is initially constant in cylinders

\begin{equation}
  T = T_0 \left(\frac{R}{R_0}\right)^q
\end{equation}

\noindent where $R=r\sin\theta$ is the cylindrical radius. Solving the condition of hydrostatic
equilibrium in the vertical direction, the density is 

\begin{equation}
  \ln \rho = \ln \rho_0   +  p\ln\left(\frac{R}{R_0}\right) + \left(1-\sin\theta\right) \left(\frac{H}{R}\right)^{-2}.
\end{equation}

\noindent We use $\rho_0=1$ and $p=-1.5$, where $p$ is the slope of the radial scaling of the midplane density.

For these profiles of density and temperature, the condition of
centrifugal equilibrium is satisfied for

\begin{equation}
  u_\phi = R\left(\ksi\varOmega_K - \varOmega_p\right)
\end{equation}

\noindent where

\begin{equation}
  \ksi^2=1 + (p+q)\left(\frac{H}{R}\right)^2 + q(1-\sin\theta).\\
\end{equation}

We use units such that 

\begin{equation}
  r_0=G(M_\star+M_P)=1
\end{equation}

\noindent and also $\rho_0$=1 as stated before. The other quantities are $c_{s0}   = 0.07$, $\gamma=  1.4$, $\mu
=2.34$, $q=-1$, and $GM_p=\xtimes{5}{-3}$ (that is,
$GM_\star=0.995$). The inclusion of radiation physics breaks the
scale-invariance of the code units. We use physical units such that 
the unit of length is the semimajor axis of Jupiter, 5.2AU, 
($\xtimes{7.785}{13}$ \ cm), and the velocity unit is thus the circular velocity at Jupiter's orbit,
$\xtimes{1.306}{6}$ cm\,s$^{-1}$. The unit of density is
$\xtimes{2}{-11}$ g\,cm$^{-3}$, as in the minimum mass solar nebula
(Hayashi 1981), and the unit of temperature is Kelvin. 

The planetary mass is increased from zero to the full mass in ten orbits, to
avoid the strong impact of introducing a high-mass planet in a
quiescent disk.

For boundary conditions in the radial direction, we use outflow for the radial
velocity, symmetric (zero-gradient) for the meridional velocity, and
constant gradient (zero second derivative) for the azimuthal
velocity, density, and entropy. 
For the meridional direction, we use zero-gradient for the radial
velocity, outflow for the meridional velocity, and constant gradient
for azimuthal velocity and entropy. For the density we maintain the
condition of vertical stratification (i.e. we extrapolate the Gaussian
structure). 

The optical depth calculated from this initial condition is plotted in
the upper panel of \fig{fig:opacity-cooling} while the 
dimensionless cooling time, $\beta=t_{\rm
  cool}\varOmega_K$, is shown in the lower panel. The expected
behaviour of nearly adiabatic in the 
midplane and nearly isothermal in the atmosphere is well reproduced. 

\section{Results}
\label{sect:results}

\begin{figure*}
  \begin{center}
    \resizebox{0.940625\columnwidth}{!}{\includegraphics{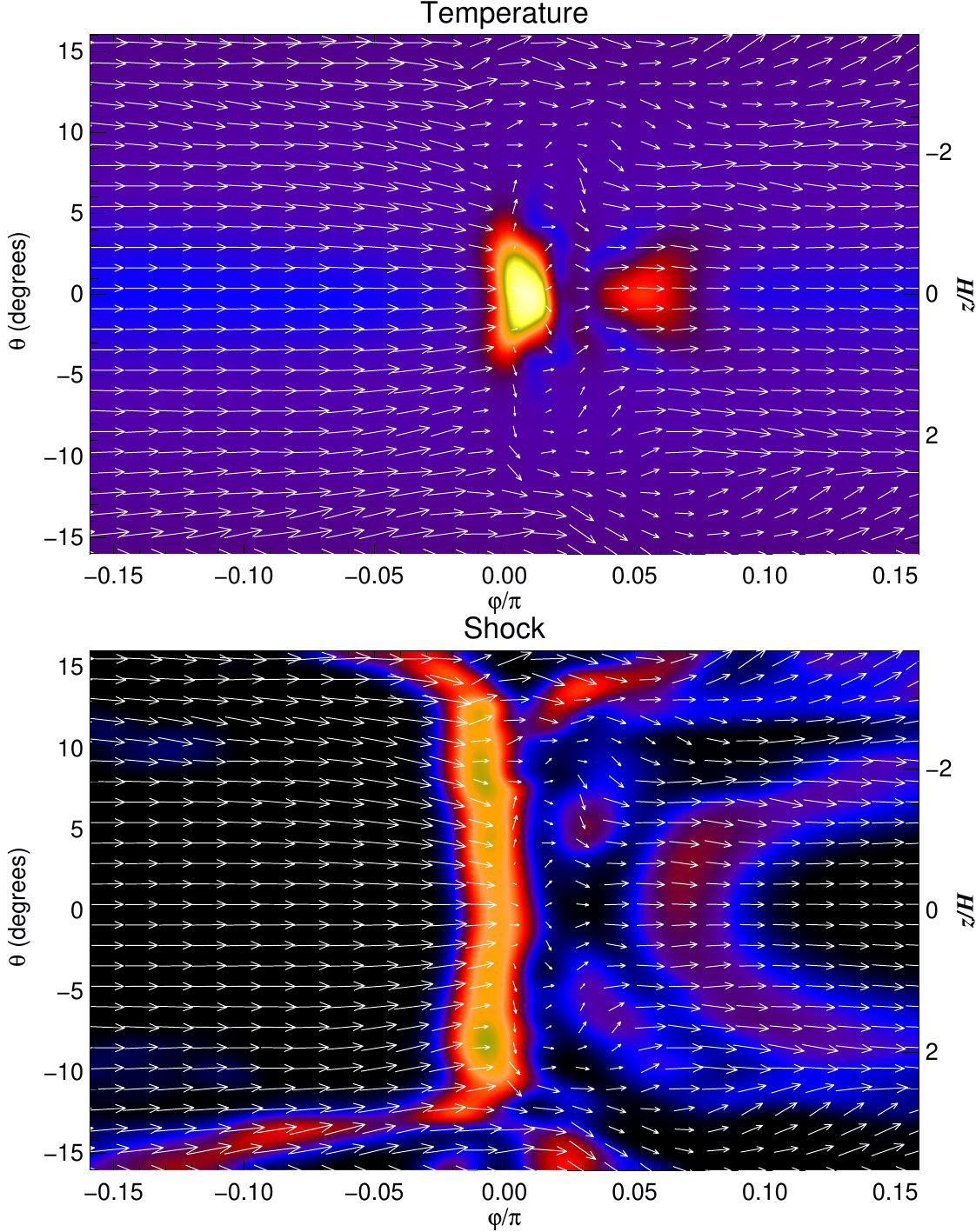}}
    \resizebox{1.05938\columnwidth}{!}{\includegraphics{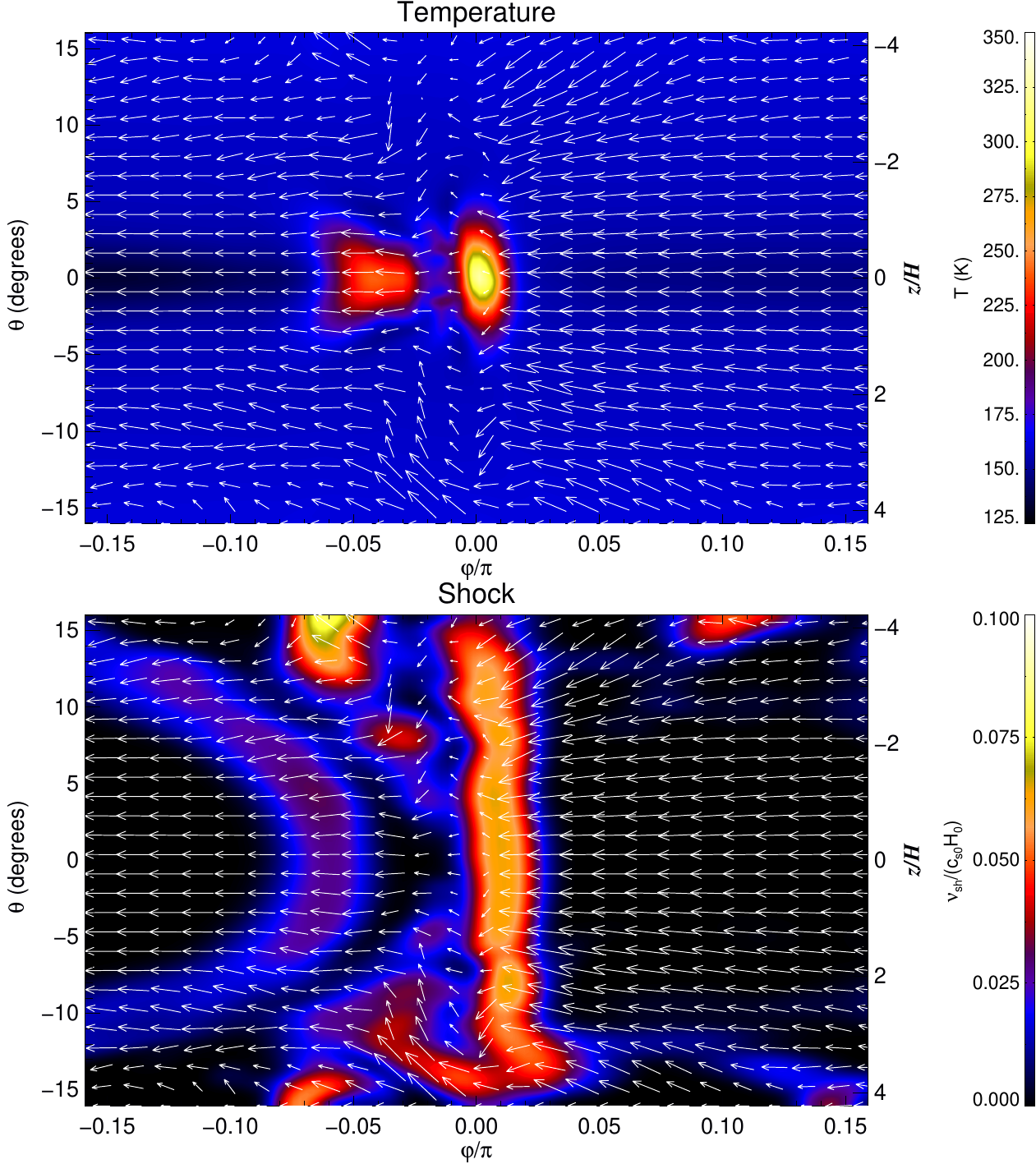}}
  \end{center}
  \caption[]{The flow structure in the $\phi-\theta$ vertical
    plane at the radial location of the Lindblad lobes interior (left
    panels) and exterior (right panels) to the planet's orbit. We show
    temperature (upper panels) and the location of shocks as traced by
    the strength of the shock viscosity (lower panels). The
    velocity field is superposed (in the reference frame of the
    planet). The flow upstream is quiescent, following Keplerian
    orbits. Upon encountering the shock, it is accelerated upwards as
    it cannot maintain hydrostatic equilibrium. The gas goes up to
    $\approx 2H$ before falling back to the midplane. The fallback is
    similar to breaking waves, generating turbulence as it propagates
    forward (\fig{fig:utht-midplane}).  The secondary bow shock is
    produced by the spiral propagation of the shock from the other
    side of the planet cutting through this radial location.}
  \label{fig:breakingwaves}
\end{figure*}

\begin{figure}
  \begin{center}
    \resizebox{\columnwidth}{!}{\includegraphics{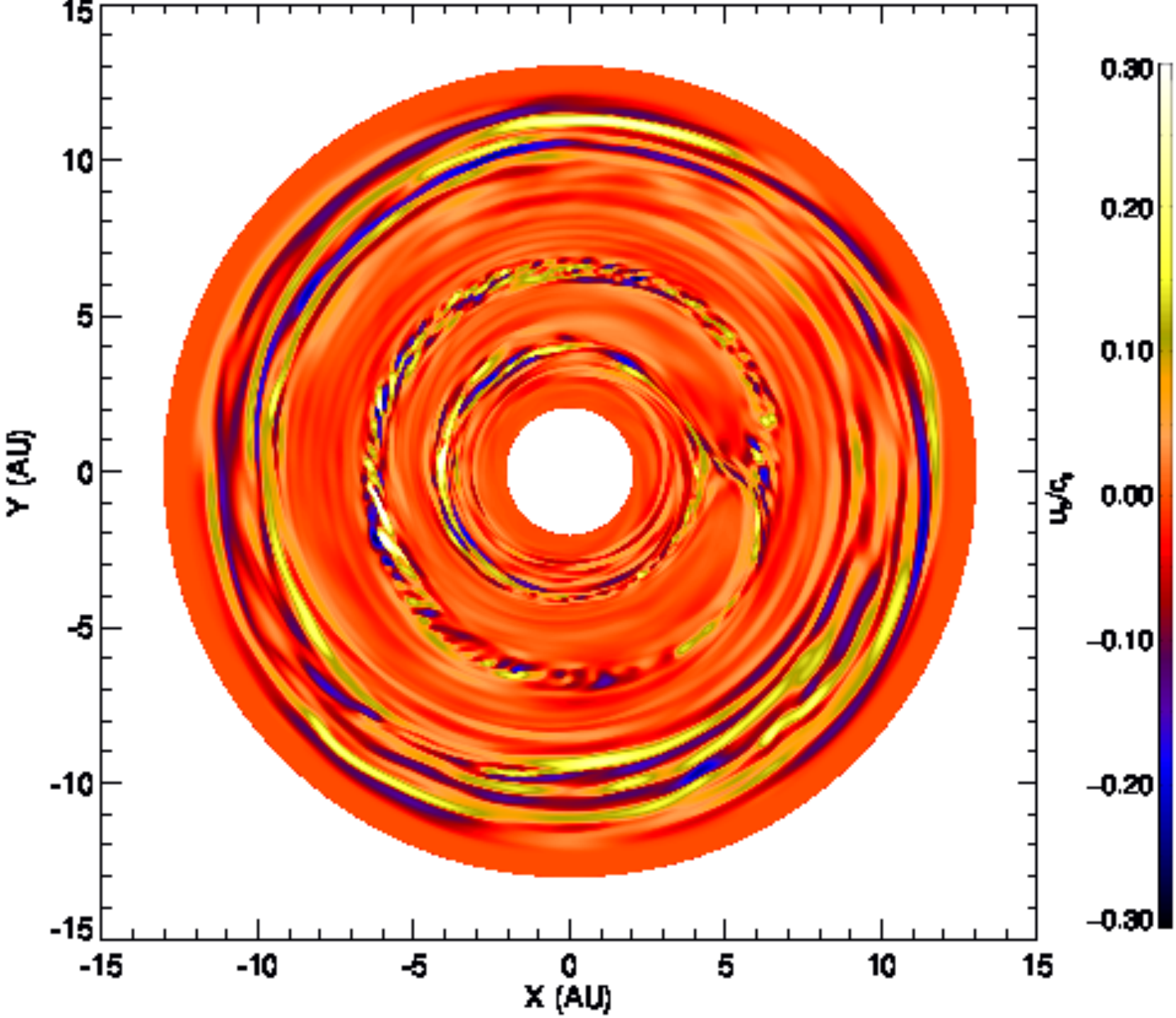}}
    \resizebox{\columnwidth}{!}{\includegraphics{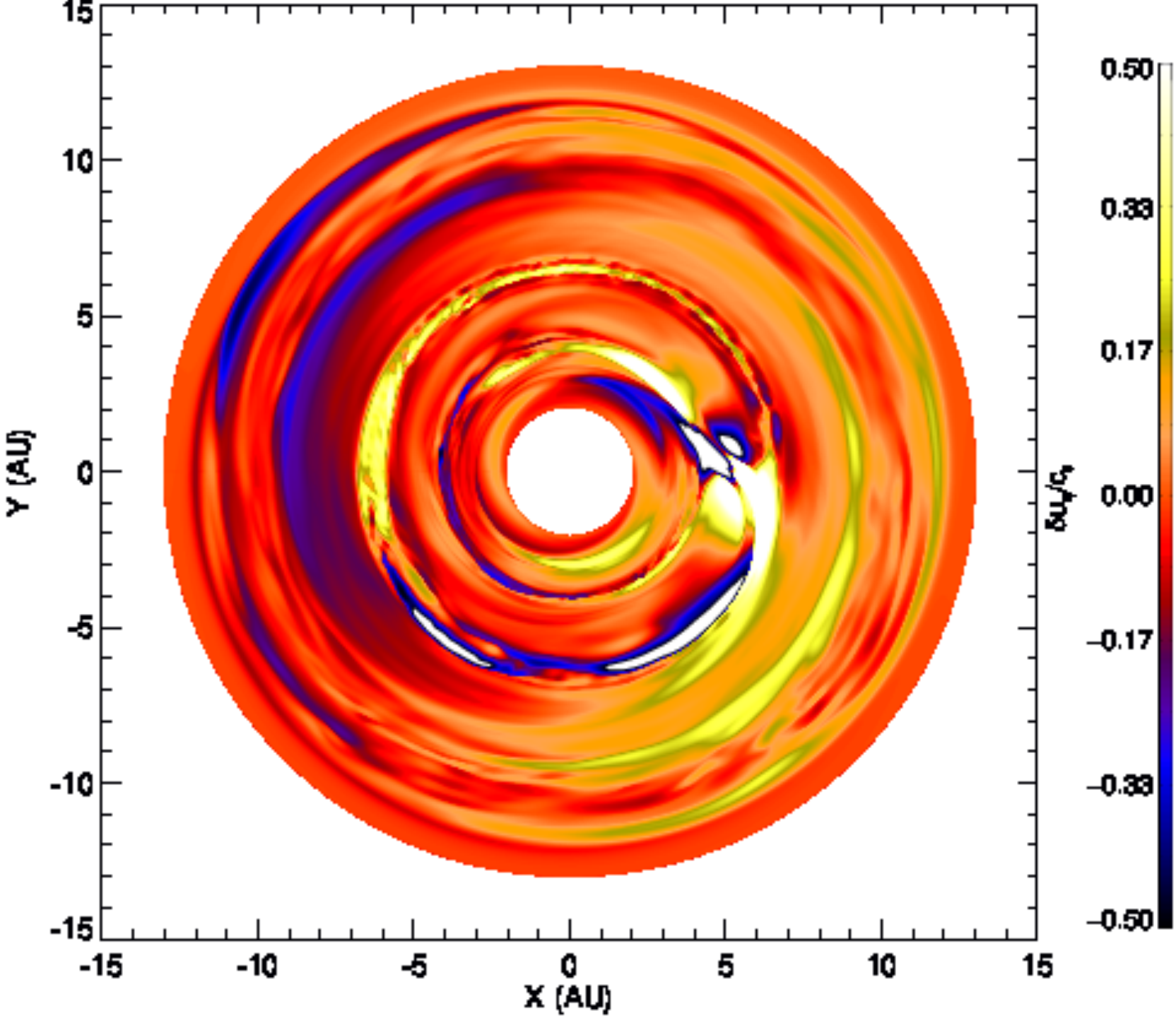}}
  \end{center}
  \caption[]{{\it Upper plot:} Vertical velocity in the midplane. The shocks at the Lindblad lobes generate turbulence
    along their orbits. The rms velocity of the turbulence is 10\% of
    the local sound speed, with maximum velocities of 40\% of the
    sound speed. {\it Lower plot:} Azimuthal velocity residuals in the
  midplane. The velocity is supersonic near the planet, with Mach
  numbers above 2. To highlight the turbulence, we plot it only up
  to Mach numbers 0.5.}
  \label{fig:utht-midplane}
\end{figure}

\begin{figure}
  \begin{center}
    \resizebox{\columnwidth}{!}{\includegraphics{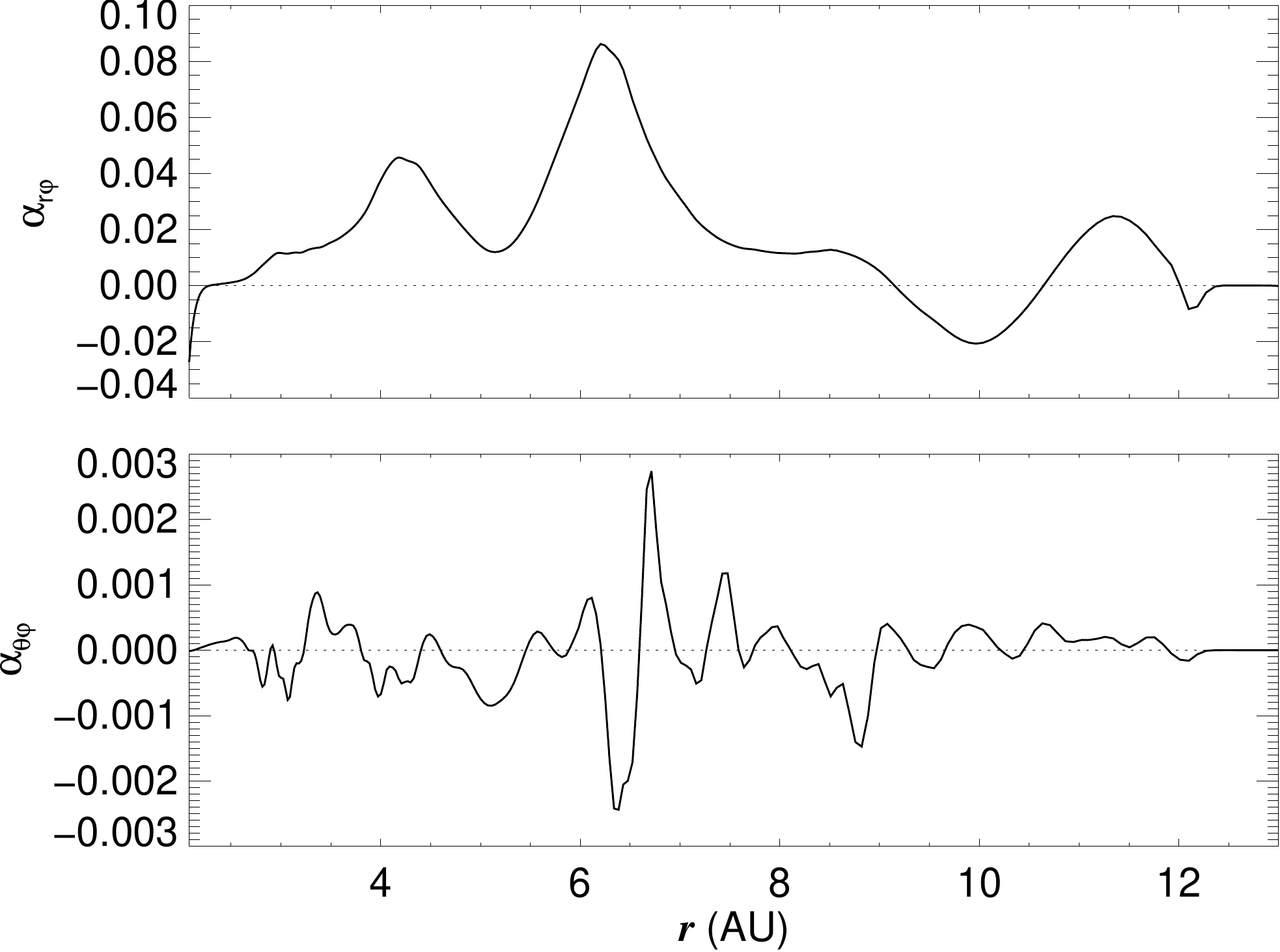}}
  \end{center}
  \caption[]{Shakura-Sunyaev (1973) $\alpha$ viscosity values leading to
    angular momentum transport in
    the radial and vertical directions (upper and lower panels,
    respectively). While the vertical transport is weak, the
    radial transport of angular momentum reaches an average value of
    $\approx0.05$ in the inner Lindblad lobe, and $\approx0.1$ in the
    outer one. This should lead to strong accretion, with
    $v_{\rm rms}\equiv\sqrt{|\alpha|}c_s \approx 0.3c_s$.}
  \label{fig:alphas}
\end{figure}

\begin{figure}
  \begin{center}
    \resizebox{\columnwidth}{!}{\includegraphics{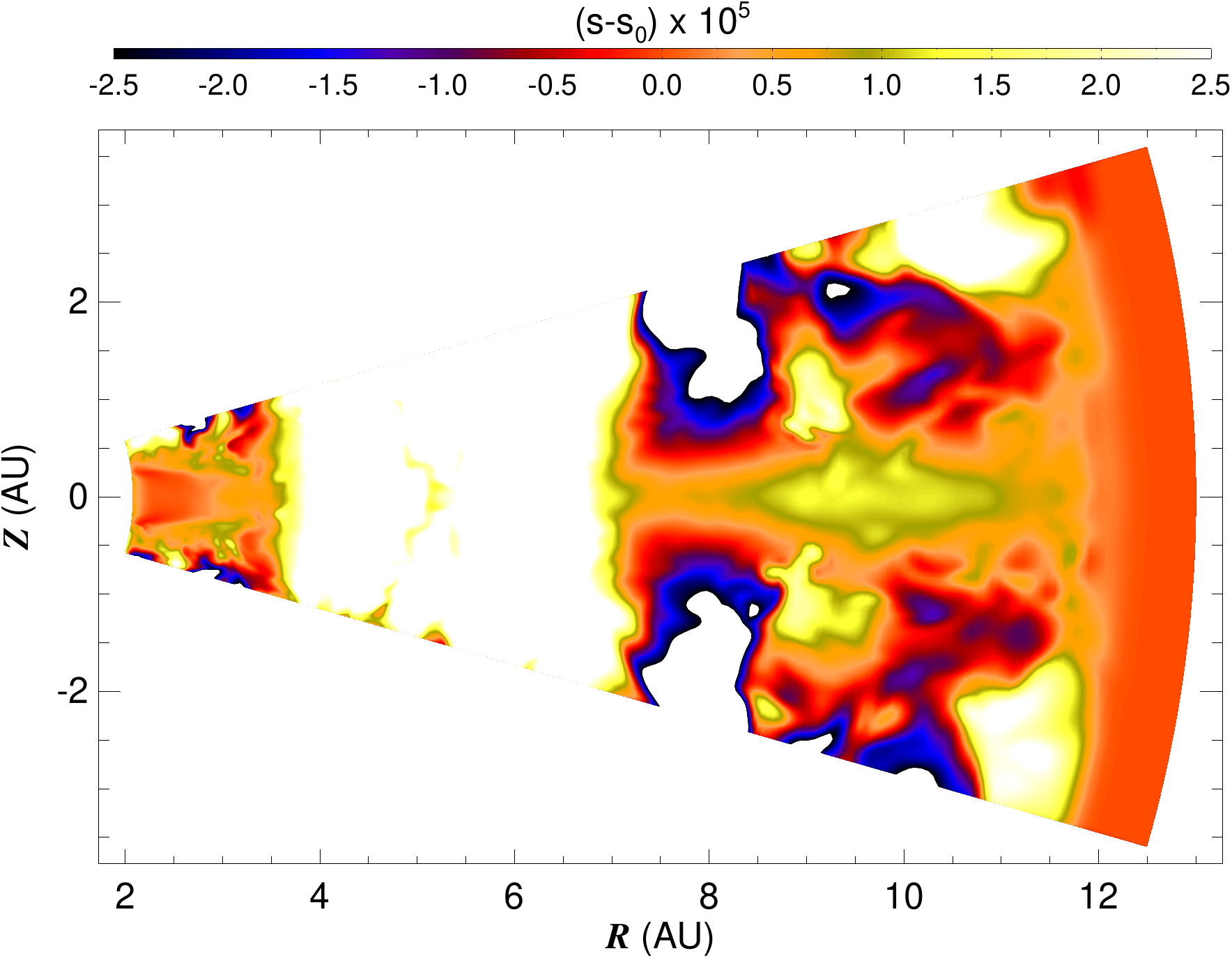}}
    \resizebox{\columnwidth}{!}{\includegraphics{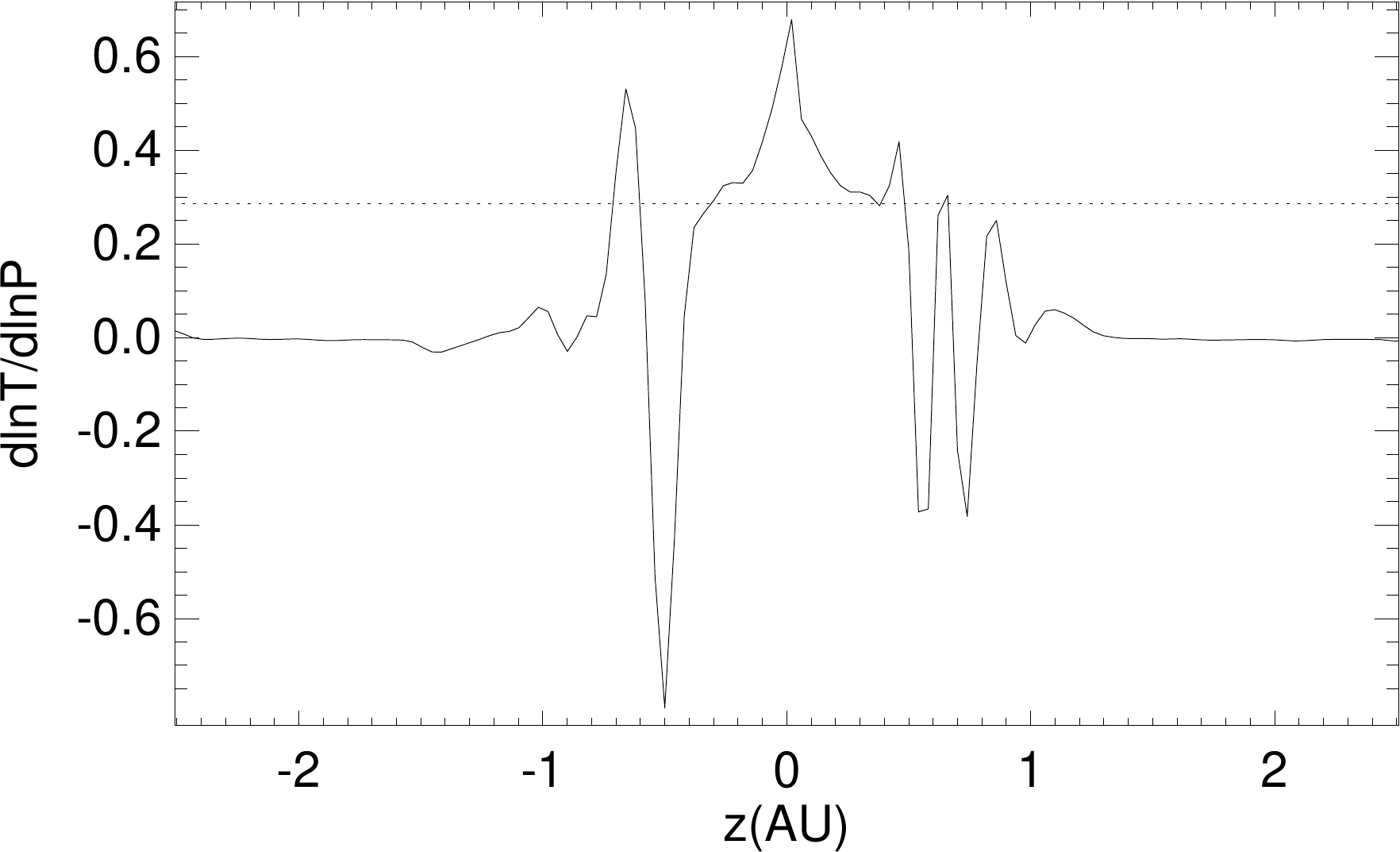}}
  \end{center}
  \caption[]{{\it Upper:} Entropy in the meridional plane that passes
    through the planet, compared to the initial
    condition. The plot is saturated, to avoid domination by the
    structures near the planet. \Fig{fig:temperature-meridional} shows structure in the
    outer disk that appears to be convection. Indeed, a blob of enhanced entropy is seen
    in the midplane, from about 8.5 to 11AU.  {\it Lower:} We check the gradient $d\ln T/d\ln P$ (solid
    line) against the adiabatic gradient (dotted line) at r=9\,AU,
    i.e, cutting through near the center of high entropy blob. The
    gradient is superadiabatic in the blob, so we conclude that indeed
    a high-mass planet generates not only turbulence in the Lindblad
    lobes but also convection far from its orbital radius along the
    generated spiral.}
  \label{fig:entropy-meridional}
\end{figure}

\begin{figure}
 \begin{center}
   \resizebox{\columnwidth}{!}{\includegraphics{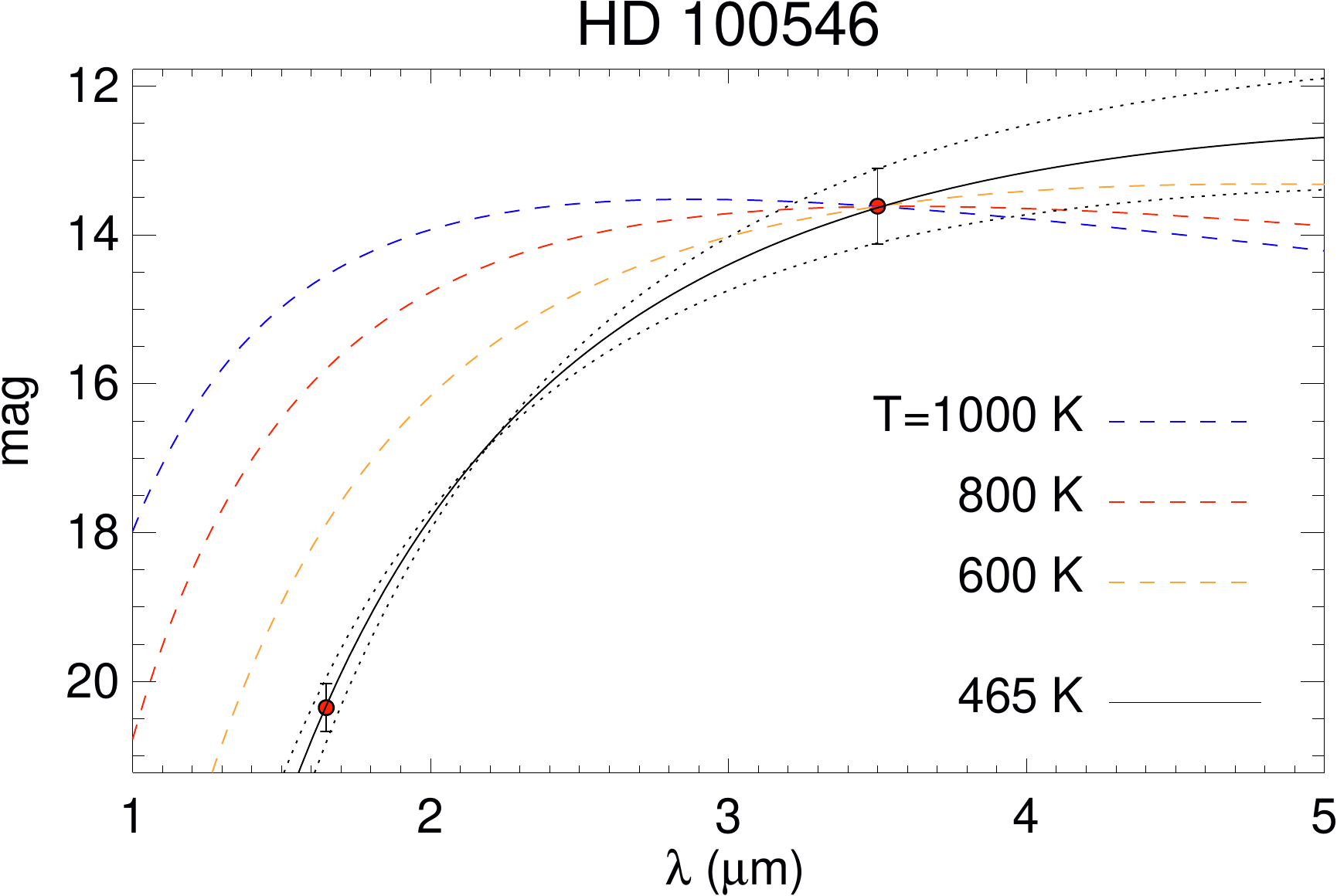}}
 \end{center}
 \caption[]{Temperature estimate for the unpolarized bright
     spiral feature in HD 100546. The data is taken from Currie et
     al. (2014), and Currie et al. (2015) for the , $L^\prime$ and
     $H$ bands, respectively (see text). The curves shown are
     blackbody curves. The best fit yields $465\pm 40$\,K; the dotted
     lines mark the $1\sigma$ standard deviation of 40\,K. The
     Planckians of 1000\,K, 800\,K and 600\,K that pass by the $L^\prime$
     data point are shown for comparison. The best fit temperature is
     remarkably close to the maximum temperature of 450\,K we find in our model.}
 \label{fig:currie}
\end{figure}

We consider a model with a 5~$M_J$ planet embedded in the disk model
described in \sect{sect:model}.  In 
\fig{fig:shocks} we show the shock structure in the disk, as traced by
the shock viscosity \eq{eq:shock-visc} after
40 orbits. (We normalize by $c_sH$ so that
the shock viscosity is dimensionless.) The left panel
shows the meridional plane that passes through the planet, while the right
panel shows the midplane. Shocks are
seen inward and outward from the planet. In the meridional plane they
form shock columns that extend well into the disk
atmosphere. In the upper atmosphere the shocks
curve slightly toward the star, as also seen in Zhu et al.\ (2015),
using a different method and cooling scheme. In the midplane the
shocks follow spiral patterns. These shocks are the primary
locations of energy dissipation, according to \eq{eq:shocktemp}. The
actual temperature increase will depend on the local value of the
cooling time $\beta$. 

\fig{fig:temperature-meridional} shows the temperature in
the meridional plane. There are high-temperature lobes in the
shock regions around the planet, and high temperatures are confined to the
more adiabatic, high-$\beta$ region around the midplane. Above the
height where $\beta\approx 1$ the 
temperature matches that of the quiescent gas. These hot lobes mark
the positions of the shock bores. In this case, the lobes are at 450\,K, while the
ambient gas is at 175\,K. 

To understand the structure of the shocks we examine in
\fig{fig:breakingwaves} the flow in the $\phi-\theta$
plane at the radial position of each lobe. The upper panel shows the
temperature, and the lower panel the shock profile. The velocity field
is superposed. The bore is seen extending almost all the way to the 
upper boundary. The flow upstream of the shock is quiescent, following
Keplerian orbits. The downstream flow shows that material is
accelerated upward, as expected from the analysis of Boley \& Durisen
(2006). The upward moving gas
  reaches approximately $2H$ before falling to the
midplane again as it proceeds downstream.

A secondary weaker bow shock forms just downstream, leading to a
smaller lobe of warm temperature. However, this bow shock does not seem to be
associated with breaking waves: it is instead the result of the
spiral feature formed by the lobe on the other side of the planet, as seen in \fig{fig:shocks}. The flow in the lobe to the right of
the planet (right panels) shows the same structure, of upward
acceleration in the vertical direction, fallback around $2H$, and secondary bow shock. 

In the upper panel of \fig{fig:utht-midplane} we plot the vertical velocity in the
midplane, showing the result of the shocks. A turbulent structure is seen
around the orbits at the radial location of the lobes that mark the
shocks. In the lower panel of \fig{fig:utht-midplane} we plot the
azimuthal velocity residual, $\delta u_\phi = u_\phi-<u_\phi>$, in the
midplane. The residual is supersonic near the planet, as expected, so
we plot it only up to Mach numbers of 0.5, to identify the
turbulence. 

We would like to quantify the level of angular momentum transport
that is caused by this turbulence at the corotation radius of the
region around 2/3$H$ where the
Lindblad resonances cluster together (which we henceforth call the
Lindblad lobe). For that we measure the alpha value (Shakura \& Sunyaev 1973), 

\begin{eqnarray}
  \alpha_{r\phi} &\equiv& \frac{\langle \rho\delta u_r\delta
    u_\phi\rangle}{\langle \rho c_s^2 \rangle}\nonumber\\
  &=&  \frac{\langle \rho u_r   u_\phi\rangle - \langle \rho u_r
      \rangle \langle  u_\phi\rangle - \langle \rho u_\phi \rangle \langle u_r\rangle + \langle \rho \rangle \langle u_r \rangle \langle u_\phi\rangle}{\langle \rho c_s^2 \rangle} \nonumber\\
\end{eqnarray}

This $\alpha_{r\phi}$ measures how much angular momentum is transported
radially. We can also measure how much angular momentum is
transported vertically (or meridionally) by computing

\begin{equation}
  \alpha_{\theta\phi} \equiv \frac{\langle \rho\delta u_\theta\delta
    u_\phi\rangle}{\langle \rho c_s^2 \rangle}
\end{equation}

We measure these quantities averaged in azimuth within $5^\circ$
above and below the midplane, which corresponds roughly to 1 scale
height each way. We then average the snapshots from 10 to 40
orbits. The result is plotted in \fig{fig:alphas}. The upper panel
shows the radial transport of angular momentum, the lower panel the
vertical transport. We see that the vertical transport is essentially
negligible, whereas the radial transport peaks in the region where the
Lindblad resonances cluster together (which we henceforth call the
Lindblad lobe)
with an average of $\alpha\approx0.05$ in the inner and $\alpha\approx0.09$
in the outer. Frequently the $\alpha$ value in the outer Lindblad lobe
exceeds 0.1, characterizing strong accretion with a characteristic velocity
$v_{\rm rms}\equiv\sqrt{|\alpha|}c_s\approx 0.3c_s$.

Some activity is also seen in the annulus around 10\,AU. Comparing
with \fig{fig:temperature-meridional}, we see this correlates with
incipient rolls forming at $r$=3.5, 8.5 and 11.5\,AU. These appear to
be buoyant expansion, radially and vertically, of the hot gas as it
propagates outwards from the spiral shock. We tentatively identify the rolls as
convection, which is also consistent with the
indications of turbulent flow seen from 3 to 8 \,AU in the midplane
plot of \fig{fig:shocks} (right panel) as well as
\fig{fig:utht-midplane}. To check that identification, we plot the entropy deviation
(with respect to the initial condition) in the upper panel of 
\fig{fig:entropy-meridional}. We saturate the plot to avoid domination
by the structures near the planet and highlight instead the variations
in the outer disk. Indeed, a blob of high entropy is seen
in the midplane, from about 8.5 to 11AU. 

To determine if this blob indeed
would lead to convection, we determine if the entropy gradient is
superadiabatic. The adiabatic gradient is 

\begin{equation}
  \left(\frac{d\ln T}{d\ln P}\right)_{\rm ad} = \frac{\gamma-1}{\gamma}.
\end{equation}

\noindent Gradients steeper than this will generate convection. We
plot in the lower panel of \fig{fig:entropy-meridional} the quantity
$d\ln T/d\ln P$ along the meridional axis at $r=9$~AU, which is near
the center of the blob. The dotted line traces the value of the
adiabatic gradient. We see that the gradient is indeed superadiabatic
in the blob. We conclude that we are seeing convection. A planet generates not only turbulence due to breaking
waves at its Lindblad lobes, but also heats material that will
propagate outward, generating convection far away from its
orbit. 

Finally, we notice that the upstream and downstream temperatures 
(175\,K and 450\,K, respectively) straddle the water ice sublimation point. 
This implies that ice particles passing through the hot shocks will sublimate.
The timescale of sublimation can be estimated as 
$\tau_{\rm subl} \sim R_{\rm p}\rho_\bullet v_{\rm vap}/P_{\rm sat}$
(Ros \& Johansen 2013), where $R_{\rm p}$ and $\rho_\bullet$ are 
the radius and internal density of the ice particles, respectively, 
$v_{\rm vap}$ is the thermal speed of water vapour, 
and $P_{\rm sat}$ is the saturation vapour pressure of water 
(for which we use the expression by Bauer et al. 1997). 
In the outer shock ($T \approx$ 350\,K), the sublimation timescale is 
$\sim 1~{\rm sec}$ for a decimeter-sized ice ball, meaning that the evaporation is instantaneous. 
Yet, the water vapour will quickly recondense downstream. 
Assuming that the sublimated ice release small silicate grains, 
condensation would mainly take place on the grains' surfaces (not on larger pebbles and planetesimals), 
because smaller particles have a larger surface-area/mass ratio. 
Therefore, the net effect is the production of a large number 
of small silicate grains coated by ice in the flow downstream. 

This sublimation process can potentially affect the size and 
spatial distribution of ice particles around the planet. 
At around 5 AU, micron-sized ice grains grow to decimeter-sized 
balls on a timescale of $\sim 1000$ yr (e.g., Brauer et al. 2008; Okuzumi et al. 2012). 
By contrast, the Lindblad lobes sweep disk gas along their orbits 
on a much shorter timescale, $\sim 2\pi r/c_{\rm s} \sim 100\,{\rm yr}$. 
Therefore, ice particles are unable to growth to macroscopic bodies near the orbits of the hot shocks. 
The destruction of macroscopic ice balls is unavoidable even if we consider the radial 
inward drift of the particles relative to the gas disk (Adachi et al. 1976; Weidenschilling 1977). 
At 5 AU, the radial drift velocity of decimeter-sized particles is $\sim 1\,{\rm AU}/ (10^3~{\rm yr})$, 
and therefore the particles are unable to cross the shock regions (of radial extent $\approx 0.5~{\rm AU}$)
without being swept by the shocks. 
Since the radial drift slows down as the particle size is decreased, 
the sublimation of radially drifting ice balls might lead to a pileup of dust at the shocks' orbits. 

\section{Discussion and Conclusions}
\label{sect:conclusions}

In paper I we showed that high-mass planets produce radial
buoyancy-induced turbulence in their parent disks. The source of energy
are the disk wakes generated by the planet at its Lindblad resonances,
which quickly steepen into shocks for massive planets. So long as the energy is
deposited more rapidly than it is dissipated, the temperature rises,
leading to instability and turbulence. We found that the turbulence extends
throughout most of the disk, and manifests itself more strongly when
the cooling is weaker. Indeed, in the adiabatic limit, without
cooling, shock dissipation continues to heat the disk until radial buoyancy
becomes the dominant motion. 

Paper I was a two-dimensional initial study, using Newton (beta)
cooling for exploration purposes. In the current paper we describe a 3D study of
the same process, with a more sophisticated approximation for treating
the radiation field in the disk. Defining a beta cooling as a function of
the local optical depth, we reproduce the behavior of adiabaticity
($\beta \gg 1$ ) in the midplane, and isothermality ($\beta \ll 1$) in
the atmosphere of a stratified disk. 

In the model presented, the
optical depth $\tau$ and the cooling time $\beta$ around the planet's
orbit are of the order of 100. For this model, we find that the turbulence continues to occur
in 3D, albeit for other reasons than in 2D. From 2D to 3D the main
difference is the extra degree of freedom given by the vertical
direction, so that as the gas is
  heated, it expands vertically, which
makes the shocks weaker (e.g. Bate et al.\ 2003). The
vertical expansion gives rise to tall shock columns, extending throughout the disk, from the midplane to the atmosphere. These were
referred to by Boley \& Durisen (2006) as shock bores. We find that
this process occurs at the radii whence the planet spiral wake is launched. The shocks produce 
hot lobes which, for our parameters, reach about 450\,K, three times
the temperature of the quiescent gas at the same orbit.

The temperature rise occurs in the midplane,
where the disk is more adiabatic, as expected. As the gas meets the shock,
it is accelerated upwards. Falling back onto the disk 
it leads to breaking waves. These waves in turn give rise to
turbulence in the full azimuthal extent around the orbital location of
each Lindblad lobe. We measure the resulting accretion stress by the
Shakura-Sunyaev parameter, finding
rapid angular momentum transport with $\alpha$ about 0.05 in the inner lobe, and $\approx0.1$ in the
outer. This means locally $v_{\rm rms}\approx 0.3c_s$. 

We also notice that as the heated gas expands outwards, radially
and vertically, it leads to structure in the meridional plane far
outward from the planet that
appears to be convection. Indeed, we find that these
features are associated with localized blobs of high
entropy around the midplane in the outer disk. Measuring the entropy
gradient, we also find these regions to be superadiabatic, confirming
susceptibility to convection. We thus predict that shocks from high-mass
planets produce three main features in disks: hot lobes at
the locations of the Lindblad resonances, strong turbulence at these
radii producing substantial accretion flows,
and convection far beyond the radius of the planet.

We compare the maximum temperature of 450\,K we find in the model with that needed to explain the
  unpolarized bright spiral feature in HD 100546. To determine the
  latter, we consider the magnitudes in the $H$ (Currie et al. 2015) and
  $L^\prime$ bands (Currie et al. 2014), and fit blackbody curves to
  these two data points. To determine the magnitude of the spiral
  feature we use the quoted magnitudes of HD 100546b in Currie et
  al. (2014, 2015), and scale the flux according to the signal-to-noise maps 
  (Fig~2 of both works). The magnitudes of HD 100546b are $13.06\pm0.51$ in the
  $L^\prime$ band, and $19.40\pm 0.32$ in the $H$ band. For the $L^\prime$ band
  the signal-to-noise is 5 in the planet and 3 in the arm. For the $H$
  band, the signal-to-noise is 6 in the planet and 2.5 in the
  arm. This yields magnitudes of $20.35$ and $13.61$ for the spiral
  arm in the $H$ and $L^\prime$ bands, respectively. We plot these
  magnitudes in \fig{fig:currie}; the data in abscissa are the centers of the 
  $H$ and $L^\prime$ bands, 1.65 and 3.5$\mu$m, respectively. The
  best fit yields $465\pm40$\,K. The 40\,K standard deviation in temperature is given by
  the temperatures that encompass the 1$\sigma$ error bars of the
  magnitudes. The Planckians at 1000\,K, 800\,K and 600\,K are shown
  for comparison. The temperature we just estimated is remarkably close to what we find in the
  model. Yet, we stress that in the model this temperature is confined
to $2H$ around the midplane, whereas in the HD 100546 observation it occurs in
the disk surface.

It is pertinent at this point to explicitly highlight a number
of limitations of our model. In our simple
approach for the cooling, radiation emitted from the hot lobes simply
disappears instead of heating the overlying material. A more complete
treatment is worthwhile, to see whether enough shock heat is transferred
to the photosphere to be observed. For future work, we intend to
  do ray-tracing for short-wavelength absorption and flux-limited
  diffusion for long-wavelength cooling, as in Flock et
  al. (2013). Another limitation is that we smooth the
planet potential at the Hill radius, for numerical
stability. This means that the planet potential is much shallower than
in reality, so we actually underestimate the magnitude of the shocks
and therefore also the temperature enhancement. Reducing the amount of smoothing of the
planet's potential would deepen the gravitational well and yield more
heating. Finally, in 
computing the optical depth we take the smaller of the upper
and lower values, which makes the midplane cooling too slow
by a factor of two. Replacing $\tau_{\rm eff}$ instead by the
inverse of the sum of the inverses of the lower
and upper value would give a better estimate. These two corrections
are in opposite directions, though, which will tend to minimize their
effect. Nevertheless, they should be checked in a future study.

\acknowledgments 

The simulations presented here were carried out using the Stampede
cluster of the Texas Advanced Computing Center (TACC) at The
University of Texas at Austin through XSEDE grant TG-AST140014.  M-MML
was partly supported by NASA grant NNX14AJ56G and the Humboldt
Foundation. We acknowledge discussions with Thayne Currie and thank
the anonymous referee for helpful comments. This work was performed in part at the Jet
  Propulsion Laboratory, California Institute of Technology. NJT was
  supported by grant 13-OSS13-0114 from the NASA Origins of the Solar
  System program.

\end{document}